\documentclass[12pt]{article}
\usepackage{amsmath}
\usepackage{amssymb}

\input{tcilatex}
\begin{document}

\begin{center}
\textbf{NAMBU-GOTO ACTION AND QUBIT THEORY}

\smallskip \ 

\textbf{IN ANY SIGNATURE AND IN HIGHER DIMENSIONS}

\smallskip \ 

H. Larragu\'{\i}vel$^{\ast }$\footnote{%
helder.larraguivel@red.cucei.udg.mx}, G. V. L\'{o}pez$^{\ast }$ \footnote{%
gulopez@udgserv.cencar.udg.mx} and J. A. Nieto$^{\star }$ \footnote{%
nieto@uas.edu.mx; janieto1@asu.edu.mx}

\smallskip \ 

$^{\ast }$\textit{Departameto de F\'{\i}sica de la Universidad de
Gudalajara, Guadalajara, M\'{e}xico.}

$^{\star }$\textit{Facultad de Ciencias F\'{\i}sico-Matem\'{a}ticas de la
Universidad Aut\'{o}noma} \textit{de Sinaloa, 80010, Culiac\'{a}n Sinaloa, M%
\'{e}xico.}

\bigskip \ 

\bigskip \ 

\textbf{Abstract}
\end{center}

We perform an extension of the relation between the Nambu-Goto action and
qubit theory. Of course, the Cayley hyperdeterminant is the key mathematical
tool in such generalization. Using the Wick rotation we find that in four
dimensions such a relation can be established no only in (2+2)-dimensions
but also in any signature. We generalize our result to a curved space-time
of (2$^{2n}$+2$^{2n}$)-dimensions and (2$^{2n+1}$+2$^{2n+1}$)-dimensions.

\bigskip \ 

\bigskip \ 

\bigskip \ 

\bigskip \ 

\bigskip \ 

\bigskip \ 

\bigskip \ 

\bigskip \ 

Keywords: Nambu-Goto action; qubit theory; general relativity

Pacs numbers: 04.60.-m, 04.65.+e, 11.15.-q, 11.30.Ly

March 17, 2016

\newpage

Some years ago, Duff [1] discovers hidden new symmetries in the Nambu-Goto
action [2]-[3]. It turns out that the key mathematical tool in such a
discovery is the Cayley hyperdeterminant [4]. In this pioneer work, however,
the target space-time turns out to have an associated $(2+2)$-signature,
corresponding to two time and two space dimensions. It was proved in Ref.
[5]-[6] that the Duff's formalism can also be generalized to $(4+4)$%
-dimensions and $(8+8)$-dimensions. Here, we shall prove that if one
introduces a Wick rotations for various coordinates then one can actually
extend the Duff's procedure to any signature in $4$-dimensions. Moreover, we
also prove that our method can be extended to curved space-time in $%
(2^{2n}+2^{2n})$-dimensions and $(2^{2n+1}+2^{2n+1})$-dimensions.

There are a number of physical reasons to be interested on these
developments, but perhaps the most important is that eventually our work may
be useful on a possible generalization of the remarkable correspondence
between black-holes and quantum information theory (see Refs. [7]-[10] and
references therein).

Let us start recalling the Duff's approach on the relation between the
Nambu-Goto action and the $(2+2)$-signature. Consider the Nambu-Goto action
[2]-[3],

\begin{equation}
S=\tint d\xi ^{2}\sqrt{\epsilon \det (\partial _{a}x^{\mu }\partial
_{b}x^{\nu }\eta _{\mu \nu })}.  \tag{1}
\end{equation}%
Here, the space-time coordinates $x^{\mu }$ are real function of two
parameters $(\tau ,\sigma )=\xi ^{a}$ and $\eta _{\mu \nu }$ is a flat
metric, determining the signature of the target space-time. Moreover, the
parameter $\epsilon $ takes the values $+1$ or $-1$, depending whether the
signature of $\eta _{\mu \nu }$ is Euclidean or Lorenziana, respectively.

It turns out that by introducing the world-sheet metric $g^{ab}$ one can
prove that (1) is equivalent to the action [11] (see also Ref. [12] and
references therein)

\begin{equation}
S=\tint d\xi ^{2}\sqrt{-\epsilon \det g}g^{ab}\partial _{a}x^{\mu }\partial
_{b}x^{\nu }\eta _{\mu \nu },  \tag{2}
\end{equation}%
which is, of course, the Polyakov action (see Ref. [12] and references
therein). In fact, from the expression

\begin{equation}
\partial _{a}x^{\mu }\partial _{b}x^{\nu }\eta _{\mu \nu }-\frac{1}{2}%
g_{ab}g^{cd}\partial _{c}x^{\mu }\partial _{d}x^{\nu }\eta _{\mu \nu }=0, 
\tag{3}
\end{equation}%
obtained by varying the action (2) with respect to $g^{ab}$, it is
straightforward to show that from (2) one obtains (1) and \textit{vise versa}%
. Hence, the actions (1) and (2) are equivalents.

It is convenient to define the induced world-sheet metric

\begin{equation}
h_{ab}\equiv \partial _{a}x^{\mu }\partial _{b}x^{\nu }\eta _{\mu \nu }. 
\tag{4}
\end{equation}%
Using this definition, the Nambu-Goto action (1) becomes

\begin{equation}
S=\tint d\xi ^{2}\sqrt{\epsilon \det (h_{ab})}.  \tag{5}
\end{equation}

It is not difficult to see that in $(2+2)$-dimensions the expression (4) can
be written as

\begin{equation}
h_{ab}=\partial _{a}x^{ij}\partial _{b}x^{kl}\varepsilon _{ik}\varepsilon
_{jl},  \tag{6}
\end{equation}%
where $x^{ij}$ denotes a the $2\times 2$- matrix

\begin{equation}
x^{ij}=\left( 
\begin{array}{cc}
x^{1}+x^{3} & x^{2}+x^{4} \\ 
-x^{2}+x^{4} & x^{1}-x^{3}%
\end{array}%
\right) .  \tag{7}
\end{equation}%
It is important to observe that (7) corresponds to the set $M(2,R)$ of any $%
2\times 2$-matrix. In fact, by introducing the fundamental base matrices

\begin{equation}
\begin{array}{ccc}
\delta ^{ij}\equiv \left( 
\begin{array}{cc}
1 & 0 \\ 
0 & 1%
\end{array}%
\right) , &  & \varepsilon ^{ij}\equiv \left( 
\begin{array}{cc}
0 & 1 \\ 
-1 & 0%
\end{array}%
\right) , \\ 
&  &  \\ 
\eta ^{ij}\equiv \left( 
\begin{array}{cc}
1 & 0 \\ 
0 & -1%
\end{array}%
\right) , &  & \lambda ^{ij}\equiv \left( 
\begin{array}{cc}
0 & 1 \\ 
1 & 0%
\end{array}%
\right) .%
\end{array}
\tag{8}
\end{equation}%
one observes that (7) can be rewritten as the linear combination

\begin{equation}
x^{ij}=x^{1}\delta ^{ij}+x^{2}\varepsilon ^{ij}+x^{3}\eta ^{ij}+x^{4}\lambda
^{ij}.  \tag{9}
\end{equation}

Let us now introduce the expression

\begin{equation}
h=\frac{1}{2!}\varepsilon ^{ab}\varepsilon ^{cd}h_{ac}h_{bd}.  \tag{10}
\end{equation}%
If one uses (4) one gets 
\begin{equation}
h=\det (h_{ab}).  \tag{11}
\end{equation}%
However, if one considers (6) one obtains

\begin{equation}
h=\mathcal{D}et(h_{ab}),  \tag{12}
\end{equation}%
where $\mathcal{D}et(h_{ab})$ denotes the Cayley hyperdeterminant of $h_{ab}$%
, namely

\begin{equation}
\mathcal{D}et(h_{ab})=\frac{1}{2!}\varepsilon ^{ab}\varepsilon
^{cd}\varepsilon _{ik}\varepsilon _{jl}\varepsilon _{mr}\varepsilon
_{ns}\partial _{a}x^{ij}\partial _{c}x^{kl}\partial _{b}x^{mn}\partial
_{d}x^{rs}.  \tag{13}
\end{equation}%
Of course, (11) and (12) imply that

\begin{equation}
\det (h_{ab})=\mathcal{D}et(h_{ab}).  \tag{14}
\end{equation}%
In turn, (14) means that in $(2+2)$-dimensions the Nambu-Goto action (5) can
also be written as

\begin{equation}
S=\tint d\xi ^{2}\sqrt{\mathcal{D}et(h_{ab})}.  \tag{15}
\end{equation}%
Note that, since in this case one is considering the $(2+2)$-signature one
must set $\epsilon =+1$ in (5).

In $(4+4)$-dimensions the key formula (6) can be generalized as

\begin{equation}
h_{ab}=\partial _{a}x^{ijm}\partial _{b}x^{kls}\varepsilon _{ik}\varepsilon
_{jl}\eta _{ms}.  \tag{16}
\end{equation}%
While in $(8+8)$-dimensions one has

\begin{equation}
h_{ab}=\partial _{a}x^{ijmn}\partial _{b}x^{klsr}\varepsilon
_{ik}\varepsilon _{jl}\varepsilon _{ms}\varepsilon _{nr}.  \tag{17}
\end{equation}%
(see Refs. [5] and [6] for details). So by considering the real variables $%
x^{i_{1}...i_{n}}$ and properly considering the matrices $\varepsilon _{ij}$
and $\eta _{ij}$ the previous formalism can be generalized to higher
dimensions. Of course, in such cases the Cayley hyperdeterminant $\mathcal{D}%
et(h_{ab})$ must be modified accordingly.

Observing (7) one wonders whether one can consider in (6) other signatures
in $4$-dimensions besides the $(2+2)$-signature. It is not difficult to see
that using the Wick rotation in any of the coordinates $x^{1}$, $x^{2}$, $%
x^{3}$ or $x^{4}$ one can modify the signature. For instance, one can
achieve the $(1+3)$-signature if one uses the prescription $x^{2}\rightarrow
ix^{2}$ in (6). This method lead us inevitable to generalize our method to a
complex structure. One simple introduce the complex matrix

\begin{equation}
z^{ij}=z^{1}\delta ^{ij}+z^{2}\varepsilon ^{ij}+z^{3}\eta ^{ij}+z^{4}\lambda
^{ij},  \tag{18}
\end{equation}%
where the variables $z^{1},z^{2},z^{3}$ and $z^{4}$ are complex numbers. The
expression (6) is generalized accordingly as [13]

\begin{equation}
h_{ab}=\partial _{a}z^{ij}\partial _{b}z^{kl}\varepsilon _{ik}\varepsilon
_{jl}.  \tag{19}
\end{equation}%
Thus, in this case, the Cayley hyperdeterminant becomes

\begin{equation}
\mathcal{D}et(h_{ab})=\frac{1}{2!}\varepsilon ^{ab}\varepsilon
^{cd}\varepsilon _{ik}\varepsilon _{jl}\varepsilon _{mr}\varepsilon
_{ns}\partial _{a}z^{ij}\partial _{b}z^{kl}\partial _{a}z^{mn}\partial
_{b}z^{rs}  \tag{20}
\end{equation}%
and consequently the Nambu-Goto action must be written using (20). Of
course, the Nambu-Goto action, or the Polyakov action, must be real and
therefore one must choose any of the coordinates $z^{1},z^{2},z^{3}$ and $%
z^{4}$ in (20) either as pure real or pure imaginary.

Similarly, the generalization to a complex structure can be made by
introducing the complex variables $z^{i_{1}...i_{n}}$ and writing

\begin{equation}
\begin{array}{c}
\mathcal{D}et(h_{ab})=\frac{1}{2!}\varepsilon ^{ab}\varepsilon
^{cd}\varepsilon _{i_{1}j_{1}}...\varepsilon _{i_{n-1}j_{n-1}}\eta
_{i_{n}j_{n}}\varepsilon _{k_{1}l_{1}...}\varepsilon
_{k_{n-1}l_{n-1}}n_{k_{n}l_{n}}\cdot \\ 
\\ 
\cdot \partial _{a}z^{i_{1}...i_{n}}\partial _{c}z^{j_{1}...j_{n}}\partial
_{b}z^{k_{1}...k_{n}}\partial _{d}z^{l_{1}...l_{n}}%
\end{array}
\tag{21}
\end{equation}%
or%
\begin{equation}
\begin{array}{c}
\mathcal{D}et(h_{ab})=\frac{1}{2!}\varepsilon ^{ab}\varepsilon
^{cd}\varepsilon _{i_{1}j_{1}}...\varepsilon _{i_{n}j_{n}}\varepsilon
_{k_{1}l_{1}...}\varepsilon _{k_{n}l_{n}}\cdot \\ 
\\ 
\cdot \partial _{a}z^{i_{1}...i_{n}}\partial _{c}z^{j_{1}...j_{n}}\partial
_{b}z^{k_{1}...k_{n}}\partial _{d}z^{l_{1}...l_{n}},%
\end{array}
\tag{22}
\end{equation}%
depending whether the signature is $(2^{2n}+2^{2n})$ or $(2^{2n+1}+2^{2n+1})$%
, respectively.

One can further generalize our procedure by considering a target curved
space-time. For this purpose let us introduce the curved space-time metric

\begin{equation}
g_{\mu \nu }=e_{\mu }^{A}e_{\nu }^{B}\eta _{AB}.  \tag{23}
\end{equation}%
Here, $e_{\mu }^{A}$ denotes a vielbein field and $\eta _{AB}$ is a flat
metric. The Polyakov action in a curved target space-time becomes

\begin{equation}
S=\tint d\xi ^{2}\sqrt{-\epsilon \det g}g^{ab}\partial _{a}x^{\mu }\partial
_{b}x^{\nu }g_{\mu \nu }.  \tag{24}
\end{equation}%
Using (23), one sees that this action can be written as

\begin{equation}
S=\tint d\xi ^{2}\sqrt{-\epsilon \det g}g^{ab}(\partial _{a}x^{\mu }e_{\mu
}^{A})(\partial _{b}x^{\nu }e_{\nu }^{B})\eta _{AB}.  \tag{25}
\end{equation}%
So, by defining the quantity

\begin{equation}
E_{a}^{A}\equiv \partial _{a}x^{\mu }e_{\mu }^{A},  \tag{26}
\end{equation}%
the action in (25) reads as

\begin{equation}
S=\tint d\xi ^{2}\sqrt{-\epsilon \det g}g^{ab}E_{a}^{A}E_{b}^{B}\eta _{AB}. 
\tag{27}
\end{equation}%
Hence, in a target space-time of $(2+2)$-dimensions one can write (27) in
the form

\begin{equation}
S=\tint d\xi ^{2}\sqrt{-\epsilon \det g}g^{ab}E_{a}^{ij}E_{b}^{kl}%
\varepsilon _{ik}\varepsilon _{jl},  \tag{28}
\end{equation}%
where

\begin{equation}
E_{a}^{ij}\equiv \partial _{a}x^{\mu }e_{\mu }^{ij}.  \tag{29}
\end{equation}%
Here, we considered the fact that one can always write

\begin{equation}
e_{\mu }^{ij}=e_{\mu }^{1}\delta ^{ij}+e_{\mu }^{2}\varepsilon ^{ij}+e_{\mu
}^{3}\eta ^{ij}+e_{\mu }^{4}\lambda ^{ij}.  \tag{30}
\end{equation}

Observe that in this development one can consider a generalization of (4)
namely

\begin{equation}
h_{ab}=E_{a}^{A}E_{b}^{B}\eta _{AB}  \tag{31}
\end{equation}%
and therefore in $(2+2)$-dimensions this expression becomes

\begin{equation}
h_{ab}=E_{a}^{ij}E_{b}^{kl}\varepsilon _{ik}\varepsilon _{jl},  \tag{32}
\end{equation}%
while in $(4+4)$-dimensions and $(8+8)$-dimensions one obtains

\begin{equation}
h_{ab}=E_{a}^{ijm}E_{b}^{klr}\varepsilon _{ik}\varepsilon _{jl}\eta _{mr} 
\tag{33}
\end{equation}%
and

\begin{equation}
h_{ab}=E_{a}^{ijmn}E_{b}^{klrs}\varepsilon _{ik}\varepsilon _{jl}\varepsilon
_{mr}\varepsilon _{ns},  \tag{34}
\end{equation}%
respectively.

At this stage, it is evident that if one wants to generalize the procedure
to any signature in a curved space-time one simply substitute in the action
(27) either

\begin{equation}
h_{ab}=\mathcal{E}_{a}^{i_{1}...i_{n}}\mathcal{E}_{b}^{j_{1}...j_{n}}%
\varepsilon _{ik}...\varepsilon _{i_{n-1}j_{n-1}}\eta _{i_{n}j_{n}}  \tag{35}
\end{equation}%
or

\begin{equation}
h_{ab}=\mathcal{E}_{a}^{i_{1}...i_{n}}\mathcal{E}_{b}^{j_{1}...j_{n}}%
\varepsilon _{ik}...\varepsilon _{i_{n-1}j_{n-1}}\varepsilon _{i_{n}j_{n}}, 
\tag{36}
\end{equation}%
depending whether the signature is $(2^{2n}+2^{2n})$ or $(2^{2n+1}+2^{2n+1})$%
, respectively. Here, we used the prescription $E_{a}^{i_{1}...i_{n}}%
\rightarrow \mathcal{E}_{a}^{i_{1}...i_{n}}$, with $\mathcal{E}%
_{a}^{i_{1}...i_{n}}$ a complex function.

In order to include $p$-branes in our formalism, one notes that the
expression (35) and (36) can still be used. In such a case, one allows the
indice $a$ in (35) and (36) to run from $0$ to $p$. Braking such kind of
indices as $a=(\hat{a}_{1},\hat{a}_{2})$ for a $3$-brane, as $a=(\hat{a}_{1},%
\hat{a}_{2},\hat{a}_{3})$, for a $5$-brane and so on one observes that (35)
and (36) can be written as

\begin{equation}
h_{\hat{a}_{1}...\hat{a}_{2}\hat{b}_{1}...\hat{b}_{2}}=\mathcal{E}_{\hat{a}%
_{1}...\hat{a}_{2}}^{i_{1}...i_{p}}\mathcal{E}_{\hat{b}_{1}...\hat{b}%
_{2}}^{j_{1}...j_{p}}\varepsilon _{ik}...\varepsilon _{i_{p-1}j_{p-1}}\eta
_{i_{p}j_{p}}  \tag{37}
\end{equation}%
or

\begin{equation}
h_{\hat{a}_{1}...\hat{a}_{2}\hat{b}_{1}...\hat{b}_{2}}=\mathcal{E}_{\hat{a}%
_{1}...\hat{a}_{2}}^{i_{1}...i_{p}}\mathcal{E}_{\hat{b}_{1}...\hat{b}%
_{2}}^{j_{1}...j_{p}}\varepsilon _{ik}...\varepsilon
_{i_{p-1}j_{p-1}}\varepsilon _{i_{p}j_{p}},  \tag{38}
\end{equation}%
respectively. The analogue of Cayley hyperdeterminant in this case will be

\begin{equation}
\begin{array}{c}
\mathcal{\hat{D}}et(h_{\hat{a}_{1}...\hat{a}_{2}\hat{b}_{1}...\hat{b}_{2}})=
\\ 
\\ 
=\varepsilon ^{\hat{a}_{1}\hat{b}_{1}}...\varepsilon ^{\hat{a}_{p}\hat{b}%
_{p}}\mathcal{E}_{\hat{a}_{1}...\hat{a}_{2}}^{i_{1}...i_{p}}\mathcal{E}_{%
\hat{b}_{1}...\hat{b}_{2}}^{j_{1}...j_{p}}\varepsilon _{ik}...\varepsilon
_{i_{p-1}j_{p-1}}\varepsilon _{i_{p}j_{p}}%
\end{array}
\tag{39}
\end{equation}%
and therefore the corresponding Nambu-Goto action becomes

\begin{equation}
S=\tint d\xi ^{p+1}\sqrt{\epsilon \mathcal{\hat{D}}et(h_{\hat{a}_{1}...\hat{a%
}_{2}\hat{b}_{1}...\hat{b}_{2}})}.  \tag{40}
\end{equation}

Summarizing, we have generalized the Duff's procedure concerning the
combination of the Nambu-Goto action and the Cayley hyperdeterminant in
target space-time of $(2+2)$-dimensions. Such a generalization first
corresponds to a curved worlds with $(2^{2n}+2^{2n})$-signature or $%
(2^{2n+1}+2^{2n+1})$-signature. Using complex structure we may be able to
extend the procedure to any signature. Further, we generalize the method to $%
p$-branes.

It turns out that these generalization may be useful in a number of physical
scenario beyond string theory and $p$-branes. In fact, since the quantity $%
z^{j_{1}...j_{n}}$ can be identified with a $n$-qubit one may be interested
in the route leading to oriented matroid theory [14] (see also Ref.
[15]-[16]). In this direction, using the phirotope concept (see Ref. [17]
and references therein), which is a complex generalization of the concept of
chirotope in oriented matroid theory, a link between super $p$-branes and
qubit theory has already been established [17]. Thus, it may be interesting
for further developments to explore the connection between the results of
the present work and supersymmetry \textit{via} the Grassmann-Pl\"{u}cker
relations (see Refs. [8]-[9] and references therein). It is worth mentioning
that such relations are natural mathematical notions in information theory
linked to $n$-qubit entanglement. Indeed, in such a case, the Hilbert space
can be broken in the form $C^{2n}=C^{L}\otimes C^{l}$ with $L=2n-1$ and $l=2$%
. This allows a geometric interpretation in terms of the complex
Grassmannian variety $Gr(L,l)$ of $2$-planes in $C^{2n}$ \textit{via} the Pl%
\"{u}cker embedding. In this context, the Pl\"{u}cker coordinates of
Grassmannians $Gr(L,l)$ are natural invariants of the theory (see Ref. [9]
for details). However, it has been mentioned in Ref. [18], and proved in
Refs. [19] and [20], that for normalized qubits the complex $1$-qubit, $2$%
-qubit and the $3$-qubit are deeply related to division algebras via the
Hopf maps, $S^{3}\overset{S^{1}}{\longrightarrow }S^{2}$, $S^{7}\overset{%
S^{3}}{\longrightarrow }S^{4}$ and $S^{15}\overset{S^{7}}{\longrightarrow }%
S^{8}$, respectively. In order to clarify the possible application of these
observations in the context of our formalism let us consider the general
complex state $\mid \psi >\in C^{2n}$, 
\begin{equation}
\mid \psi
>=\tsum%
\limits_{i_{1}i_{2}...i_{n}=0}^{1}z^{i_{1}i_{2}...i_{n}}|i_{1}i_{2}...i_{n}>,
\tag{41}
\end{equation}%
where $|i_{1}i_{2}...i_{n}>=|i_{1}>\otimes |i_{2}>\otimes ...\otimes |i_{n}>$
correspond to a standard basis of the $n$-qubit. It is interesting to make
the following observations. First, let us denote a $n$-rebit system (real $n$%
-qubit ) by $x^{i_{1}i_{2}...i_{n}}$. So, one finds that a $3$-rebit and $4$%
-rebit have $8$ and $16$ real degrees of freedom, respectively. Thus, one
learns that the $4$-rebit can be associated with the $16$ degrees of freedom
of a $3$-qubit. It turns out that this is the kind of embedding discussed in
Ref. [9]. In this context, one sees that in the Nambu-Goto context one may
consider the $16$-dimensions target space-time as the maximum dimension
required by division algebras via the Hopf map $S^{15}\overset{S^{7}}{%
\longrightarrow }S^{8}$.

\bigskip \ 

\begin{center}
\textbf{Acknowledgments}

\smallskip \ 
\end{center}

J. A. Nieto would like to thank to P. A. Nieto for helpful comments. This
work was partially supported by PROFAPI/2007 and PIFI 3.3.

\bigskip \


\begin{thebibliography}{99}
\bibitem{1} M. J. Duff, Phys. Lett. B \textbf{641} (2006) 335;
hep-th/0602160.

\bibitem{2} Y. Nambu, Lectures a the Copenhagen Symposium, 1970, unpublished.

\bibitem{3} T. Goto, Progr. Theor. Phys. \textbf{46} (1971) 1560.

\bibitem{4} A. Cayley, J. Camb. Math. \textbf{4} (1845) 193.

\bibitem{5} J. A. Nieto, Phys. Lett. B \textbf{692} (2010) 43;
arXiv:1004.5372 [hep-th].

\bibitem{6} J. A. Nieto, Phys. Lett. B \textbf{718} (2013) 1543;
arXiv:1210.0928 [hep-th].

\bibitem{7} L. Borsten, D. Dahanayake, M. J. Duff, H. Ebrahim and W. Rubens,
Phys. Rept. \textbf{471} (2009) 113; arXiv: hep-th/0809.4685.

\bibitem{8} P. Levay, Phys. Rev. D \textbf{74} (2006) 024030; arXiv:
hep-th/0603136.

\bibitem{9} P. Levay, J. Phys. A \textbf{38} (2005) 9075.

\bibitem{10} P. Levay, Phys. Rev. D \textbf{82} (2010) 026002;
arXiv:1004.2346 [hep-th].

\bibitem{11} A. Polyakov Phys. Lett. B \textbf{103} (1981) 207.

\bibitem{12} M. Green, J. Schwarz, E. Witten, \textit{Superstring Theory}
(Cambridge U. Press, Cambridge,UK, 1987).

\bibitem{13} H. Larragu\'{\i}vel, G. V. L\'{o}pez and J. A. Nieto, work in
progress (2015); Bachelor Thesis "Antropic Gravity, Black-Holes and Q-bits",
H. Larragu\'{\i}vel, Physics Deparment, Guadalajara University, December
(2015).

\bibitem{14} A. Bj\"{o}rner, M. Las Vergnas, B. Sturmfels, N. White and G.
M. Ziegler, Oriented Matroids, (Cambridge University Press, Cambridge, 1993).

\bibitem{15} J. A. Nieto, Adv. Theor. Math. Phys. \textbf{10} (2006) 747;
hep-th/0506106.

\bibitem{16} J. A. Nieto, Adv. Theor. Math. Phys. \textbf{8} (2004)
177;hep-th/0310071.

\bibitem{17} J. A. Nieto, Nucl. Phys. B \textbf{883} (2014) 350;
arXiv:1402.6998 [hep-th].

\bibitem{18} R. Mosseri and R. Dandoloff, J. Phys. A: Math. Gen. \textbf{34}%
, (2001) 10243.

\bibitem{19} R. Mosseri, \textquotedblleft Two and Three Qubits Geometry and
Hopf Fibrations\textquotedblright ; arXiv:quant-ph/0310053.

\bibitem{20} B. A. Bernevig and H. D. Chen, J. Phys. A; Math. Gen. \textbf{36%
}, (2003) 8325.
\end{thebibliography}
\end{document}